\documentclass[floats,floatfix,showpacs,amssymb,amsmath,prd,twocolumn,superscriptaddress,nofootinbib, aps]{revtex4-2}
\usepackage[utf8]{inputenc}
\usepackage{hyperref, amsmath, amsfonts}
\usepackage{graphicx}
\usepackage{aas_macros}
\hypersetup{colorlinks,bookmarksnumbered,
citecolor=cyan, urlcolor=magenta}

\providecommand{\adsurl}[1]{\href{#1}{ADS}}
\usepackage{natbib}
\usepackage{booktabs}
\usepackage{ulem}
\usepackage{subcaption}

\usepackage[usenames, dvipsnames]{xcolor}

\begin{document}
\preprint{APS/123-QED}

\title{Detecting dark matter subhalos with the Nancy Grace Roman Space Telescope}
\author{Kris Pardo}
\email{kpardo@caltech.edu}
\affiliation{Jet Propulsion Laboratory, California Institute of Technology, Pasadena, CA 91101, USA}
\author{Olivier Dor\'e}
\email{olivier.p.dore@jpl.nasa.gov}
\affiliation{Jet Propulsion Laboratory, California Institute of Technology, Pasadena, CA 91101, USA}
\affiliation{California Institute of Technology, Pasadena, CA 91125, USA}
\date{\today}
\begin{abstract}
The dark matter subhalo mass function is a promising way of distinguishing between dark matter models. While cold dark matter predicts halos down to Earth-sized masses, other dark matter models typically predict a cut-off in the subhalo mass function. Thus a definitive detection or limits on the existence of subhalos at small masses can give us insight into the nature of dark matter. If these subhalos exist in the Milky Way, they would produce weak lensing signatures, such as modified apparent positions, on background stars. These signatures would generate correlations in the apparent velocities and accelerations of these stars, which could be observable given sufficient astrometric resolution and cadence. The Nancy Grace Roman Space Telescope's Exoplanet Microlensing Survey will be perfectly suited to measuring the acceleration signatures of these halos. Here we forward model these acceleration signatures and explore the Roman Space Telescope's future constraints on lens profiles and populations. We find that the Roman Space Telescope could place competitive bounds on point source, Gaussian, and Navarro-Frenk-White (NFW) profile lenses that are complementary to other proposed methods. In particular, it could place 95\% upper limits on the NFW concentration, $c_{200} < 10^{2.5}$. We discuss possible systematic effects that could hinder these efforts, but argue they should not prevent the Roman Space Telescope from placing strong limits. We also discuss further analysis methods for improving these constraints.
\end{abstract}
\maketitle

\section{Introduction}
Although we have known about dark matter (DM) for many decades \citep{Zwicky1933, Rubin1978, Rubin1980}, we have yet to find a particle responsible for its effects. The promise of a weakly interacting particle has so far eluded direct detection efforts \citep{Aprile2018, Akerib2017, Cui2017, PandaX}. Many other candidate particles have attracted considerable attention (e.g., axions and self-interacting DM); however we have not seen concrete evidence of their unique signatures.

While we have observed many DM signatures, one feature is both universal to all galaxies and discerning of different particle models: the subhalo mass function, which describes the distribution of sub-galactic DM halos within galaxies. From hierarchical structure formation, we expect smaller halos to exist within larger ones \citep{Ghigna1998, Klypin1999}. The exact properties of the DM (e.g., whether it is warm or cold, or whether it has any interactions with other particles) define the shape of the subhalo mass function  and the profiles of these halos. The cold dark matter (CDM) model predicts halos down to Earth-sized masses \citep{Green2005, Bringmann2009}. However, many other models have subhalo mass function cutoffs at higher scales. Thus, if we are able to find evidence for subhalos at small masses, we can rule out certain properties of DM \citep[see Ref.][for a review of these issues]{Bullock2017}.

For subhalos with masses $\lesssim 10^{10}~\rm{M}_\odot$, star formation is heavily suppressed by photoheating and tidal interactions with host galaxies \citep{Efstathiou1992, Hoeft2006, Okamoto2008, Noh2014}. Thus, for the smallest structures, the most reliable observation mechanism is through gravitational interactions. Microlensing has placed stringent constraints on very compact lenses, such as Massive Compact Halo Objects (MACHOs) \citep{Paczynski1986,Alcock2001,Tisserand2007}. Recently, stellar streams have also emerged as a way of constraining dense halos \citep[e.g.,][]{Bonaca2019}. For more diffuse halos, strong gravitational lensing has produced interesting constraints \citep[e.g.,][]{Nierenberg2017, Hezaveh2016, Ostdiek2020, Alexander2020}.

Subhalos could also produce astrometric signatures on stars. Initial studies focused on microlensing signatures \citep{Erickcek2011, Li2012}, which are not likely to be observable unless DM subhalos are especially compact. More recently, Ref.~\cite{vT2018} explored the weak lensing\footnote{By weak lensing, we mean that multiple images of the sources are not produced.} signatures that DM subhalos would induce on stars within the Milky Way. These vary depending on the types of halos (e.g., point source or diffuse). However, all are observed through changes in apparent stellar velocities or accelerations. Follow-up work calculated the power spectrum of these lensing signatures and forecasted interesting constraints from current and future missions \citep{Mishra-Sharma2020}. More recent work has also considered the use of machine learning to look for these signatures \citep{Vattis2020, Mishra-Sharma2021}.

The Nancy Grace Roman Space Telescope is NASA's next flagship mission after JWST, and it will have the necessary astrometric resolution to look for these weak lensing signatures \citep{Spergel2013, Akeson2019, Mishra-Sharma2020}. In particular, the Exoplanet Microlensing (EML) survey would have the precision necessary to properly probe this signal. The EML survey should continuously monitor an approximately 2 deg$^2$ patch near the galactic center for six 72-day periods. The 15 minute cadence over these fields should produce over 40,000 exposures of each star in the region \citep{Gaudi2019, Penny2019, Sanderson2019}. Combined with Roman's relative astrometric accuracy of 1.1 mas per exposure, this would create incredibly precise acceleration maps of the $10^8$ stars that the EML survey will monitor.

\begin{figure*}[!htb]
    \centering
    \includegraphics{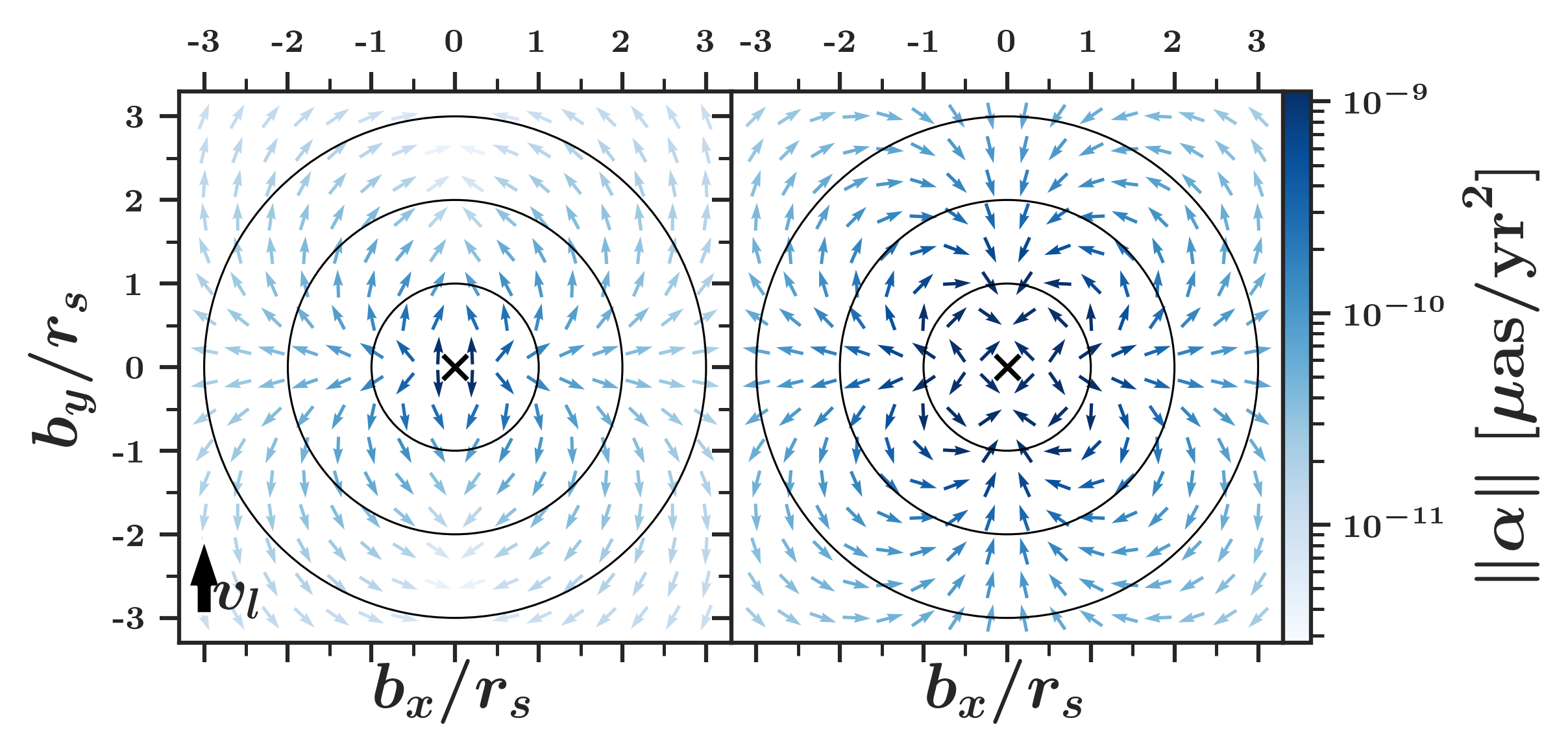}
    \caption{Example lensing-induced apparent accelerations assuming different mass profiles. The lens is located at the black `X' at the center of each plot and it is assumed to be moving in the positive $y$-direction. Each blue arrow shows the direction of the induced acceleration and its color shows the magnitude. The black circles demarcate the NFW scale radius. {\it Left:} NFW lens. {\it Right:} Point-source lens. The differences in the signals are due to the dependence on derivatives of the mass profile with respect to the impact parameter (see text below).}
    \label{fig:template}
\end{figure*}

In this paper, we forward model the constraints that the Nancy Grace Roman Space Telescope's EML Survey would place on DM subhalos in the Milky Way. In Section~\ref{sec:theory} we summarize the lensing signature we expect. In Section~\ref{sec:methods} we discuss our forward modeling approach. In Section~\ref{sec:results} we present our results, discuss possible systematics, and identify avenues for improved constraints.

The code used in this paper is publicly available at: \url{https://github.com/kpardo/dmsl}.

\section{Theory}\label{sec:theory}
In this section, we provide a summary of the lensing signatures expected by DM substructure. For a more detailed discussion of these signatures, see Ref.~\cite{vT2018}.

The apparent change in a star's position caused by one lens is:
\begin{equation}
    \Delta \vec{\theta} = -\left(1-\frac{D_l}{D_s}\right) \frac{4GM_l}{c^2b} \hat{b} \; ,
\end{equation}
where $c$ is the speed of light, $b$ is the impact parameter on the sky, $D_l$ is the distance to the lens, $D_s$ is the distance to the star, and $M_l$ is the lensing mass within a cylinder of radius $b$ and height $D_s$. From now on, we take the distant-source limit: $D_s \gg D_l$.

Then, the apparent velocity change is \citep{vT2018, Mishra-Sharma2020}:
\begin{equation}
    \vec{\mu} \equiv \Delta \dot{\vec{\theta}} = 4\frac{G}{c^2}\left[\frac{M_l}{b^2} \left(2(\hat{b}\cdot \vec{v})\hat{b} - \vec{v}\right) - \frac{M_l'}{b} (\hat{b}\cdot \vec{v})\hat{b}\right]\; ,
\end{equation}
where $\dot{} \equiv \frac{d}{dt}$, $\vec{v} \equiv \dot{\vec{b}}$, and $' \equiv \frac{d}{db}$. The apparent acceleration due to a DM lens is then:
\begin{equation}\label{eqn-alphal}
    \vec{\alpha} \equiv \Delta \ddot{\vec{\theta}} = 4\frac{G}{c^2}\left[\frac{M_l}{b^3} A(b) + \frac{M_l'}{b^2} B(b) + \frac{M_l''}{b}C(b)\right] \; ,
\end{equation}
where:
\begin{eqnarray}
   A(b) &=& -8(\hat{b}\cdot \vec{v})^2\hat{b}+ 2b(\hat{b}\cdot \dot{\vec{v}})\hat{b} + 2v^2\hat{b}
   \nonumber \\
   & &+ 4(\hat{b}\cdot \vec{v})\vec{v} - b\dot{\vec{v}}\\
   B(b) &=&  5(\hat{b}\cdot \vec{v})^2\hat{b} - 2(\hat{b}\cdot \vec{v})\vec{v}
    -b(\hat{b}\cdot \dot{\vec{v}})\hat{b} - v^2\hat{b}\\
   C(b) &=&  -(\hat{b}\cdot \vec{v})^2 \hat{b}\; .
\end{eqnarray}
For the rest of this work, we set $\dot{\vec{v}} = 0$. We also ignore any changes to the direction of $\vec{b}$.

Note the very strong dependence on the impact parameter here, $\vec{\alpha} \propto 1/b^3$. This will be important to our statistical formalism. 

Examples of the lensing signature for two different mass profiles (NFW and point-source) are given in Fig.~\ref{fig:template}. The differences in the signatures are due to the dependence on the mass profile derivatives in Equation~\ref{eqn-alphal}. For an NFW profile, all of the mass-impact parameter ratios ($M_l/b^3$, $M_l'/b^2$, and $M_l''/b$) are approximately the same order of magnitude, with $M_l''/b$ having the opposite sign as the other terms (for $b/r_s > 1$). The resulting simplifications give a dipole. For the point source case, $M_l'=M_l'' = 0$. This allows the quadrupole moment to dominate. These behaviors are easiest to see when using a vector spherical harmonic decomposition \citep[see Ref.][for some discussion of this topic]{Mishra-Sharma2020}.

\section{Methods}\label{sec:methods}
In general, we expect there to be some population of lenses that change the apparent accelerations of stars. Each star can be lensed by multiple lenses and the probability of being lensed increases with the number of lenses. Thus, the lens signal for any one star will rely on $M_l$, $v_l$, and $b$, as well as the total number of lenses, $N_l$. Here we forward model the expected distribution of acceleration signals induced by DM weak lensing. We then perform a Markov Chain Monte Carlo (MCMC) over different lens parameters to find the constraints the Roman Space Telescope would achieve. In the following, we first give the details of our statistical model and then describe our sampling efforts.

\subsection{Statistical Formalism}
Upon a first attempt at this problem, we could try to fully forward model the system. This would involve placing the lenses at random with each sample of the MCMC. However, this approach quickly becomes intractable without considerable computational resources. It requires many samples to fully probe the impact parameter distribution at any fixed $N_l$, nevermind when also sampling $M_l$, which sets $N_l$. In addition, this forward modeling would be very sensitive to the number density of simulated stars in the field of view (FOV). To achieve the correct result, a run with the same number as the true FOV would be required. Then the distances from these $10^8$ stars to the $N_l$ lenses would need to be computed at each sample. In this paper, we instead opt for a semi-analytic approach, which we find reasonably accurate and more computationally efficient. Rather than sampling lens positions, we find the distribution of expected impact parameters given the number of lenses in the field. For each star in the field, we then sample from this distribution to find its impact parameter to the nearest lens. In the following, we explain this process in more detail.

Equation~\ref{eqn-alphal} gives the apparent acceleration of 1 star from 1 lens. We first consider the likelihood of 1 star's given apparent acceleration due to a population of lenses. We will then generalize to $N_\star$ stars. Throughout, we assume that all lenses have the same mass and that their mass profiles are set solely by this mass.

First, let us consider the apparent acceleration of 1 star from a population of $N_l$ lenses with the same mass, $M_l$. We assume that we are in the weak lensing regime. Then the total acceleration from these $N_l$ lenses is just the vector sum of the signal from each lens:
\begin{equation}
    \vec{\alpha}_l = \sum_{i=1}^{N_l} \vec{\alpha}_{l,i}(M_l, \vec{v}_{l,i}, \vec{b}_{i}) \; ,
\end{equation}
where $\vec{\alpha}_{l,i}$ is given by Equation~\ref{eqn-alphal}.

In general, the closest lens will dominate (since Equation~\ref{eqn-alphal} has such a strong dependence on $b$). Thus, we now only consider the lensing signature from the closest lens to each star. The impact parameter, $b$, is then taken to always be the minimum impact parameter. Then the above equation simplifies to $\vec{\alpha}_l \approx \vec{\alpha}_{l}(M_l, \vec{v}_{l}, \vec{b}_{\rm{min}})$.

We assume that the number of lenses is uniquely determined by the mass.
Specifically, we assume that all of the dark matter is contained in halos of one mass, $M_l$. Then, to conserve the local dark matter density, the number of lenses for a given lens mass is:
\begin{equation}\label{eqn-nl}
    N_l = \frac{\rho_{\rm{DM}}A_{\rm{FOV}} d_{\rm{max}}}{3 M_l} \; ,
\end{equation}
where $\rho_{\rm{DM}}$ is the local density of dark matter, $A_{FOV}$ is the area covered by the FOV, and $d_{\rm{max}}$ is the maximum distance of the lenses. This assumes that the volume probed by the telescope is a square pyramid, which is a fine assumption for small FOVs (in other words, we take the flat sky approximation). We always assume at least $1$ lens within the volume. Note that we do not realistically expect lenses within the FOV for the very largest masses here; however, quantifying this involves an assumed subhalo mass distribution. We leave this to future work.

The number of lenses, $N_l$, then sets the impact parameter distribution as follows. First, we assume that the distribution of lenses is isotropic. This allows us to split $\vec{b}$ into an angular portion, $b_{\theta}$, which will just have a flat prior on $[0, 2\pi)$, and a magnitude portion, $b \equiv \| \vec{b} \|$. We now also split $b$ into an angular distance on the sky, $\beta$, and a distance along the line of sight, $d_l$. Specifically, we assume the small angle approximation and define $b = \beta \times d_l$. Given the FOV of the Roman Space Telescope (0.28~deg$^2$), the small angle approximation is valid. We assume that the distribution of $\beta$ is independent of the distribution of $d_l$. The distances to the lenses are given by simply assuming that they follow the dark matter density profile of the Milky Way. In particular, we assume the lenses' radial positions follow a Navarro, Frenk \& White \citep[NFW;][]{Navarro1997} profile:
\begin{equation}\label{eqn:mwdens}
    \rho(r) = \frac{4\rho_s}{\frac{r}{r_s}(1+\frac{r}{r_s})^2} \; ,
\end{equation}
where $r$ is the distance from the MW center, $r_s = 18~\rm{kpc}$ is the scale radius of the MW halo, and $\rho_s = 0.003~M_{\odot}/\rm{pc}^3$ is the density at that radius \citep[i.e., we assume the same parameters as Ref.][]{Mondino2020}. We also assume that all lenses are within $1~\rm{kpc}$ of the Sun. Ref.~\cite{Mishra-Sharma2020} found that most of the power from DM lenses is constrained within this radius. This constraint also allows us to continue assuming the distant source limit.

\begin{figure*}[!thb]
    \begin{subfigure}{0.49\textwidth}
    \centering
    \includegraphics[width=0.9\linewidth]{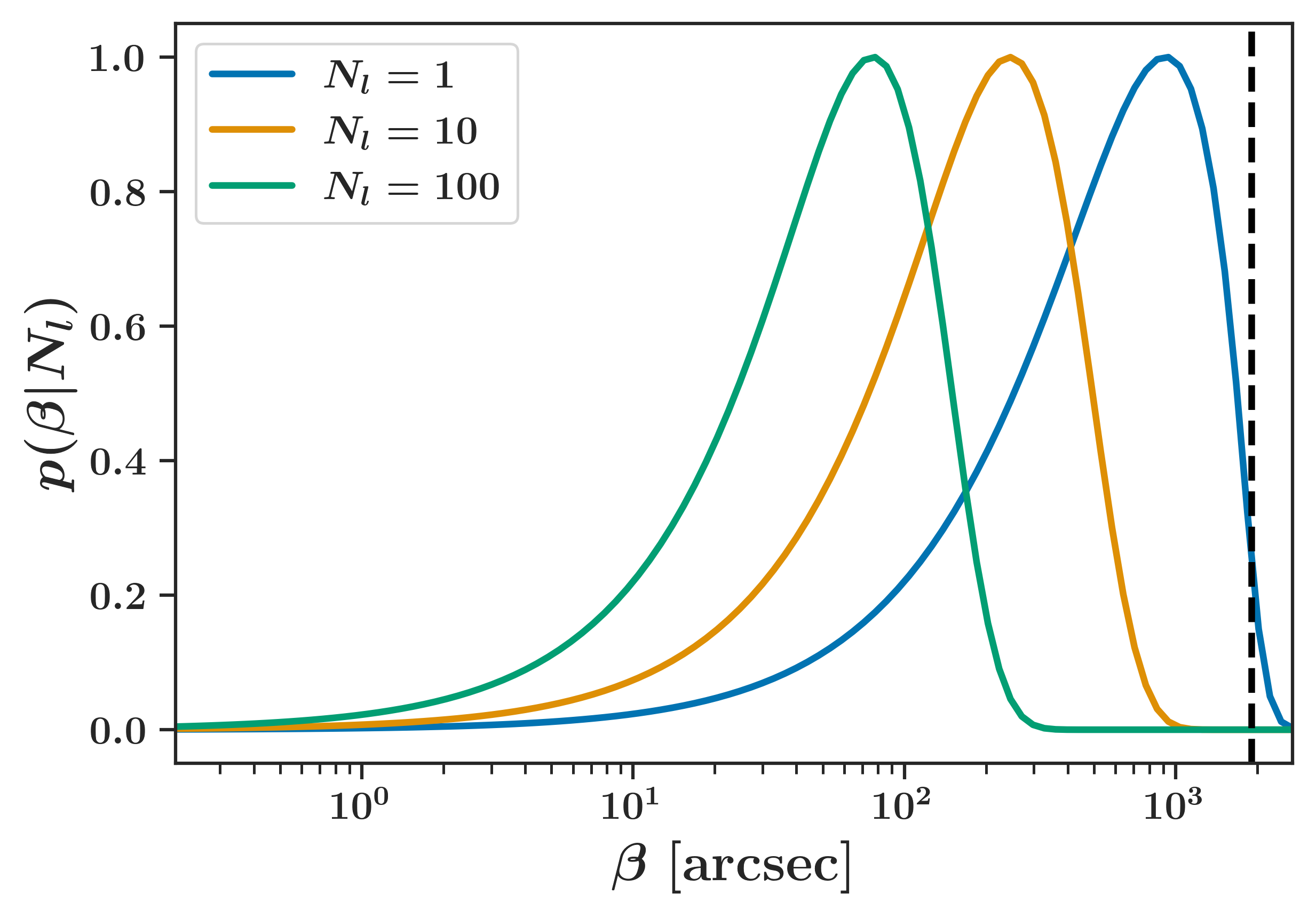}
    \caption{}
    \label{fig:impact_prior}
    \end{subfigure}
    \begin{subfigure}{0.49\textwidth}
    \centering
    \includegraphics[width=0.9\linewidth]{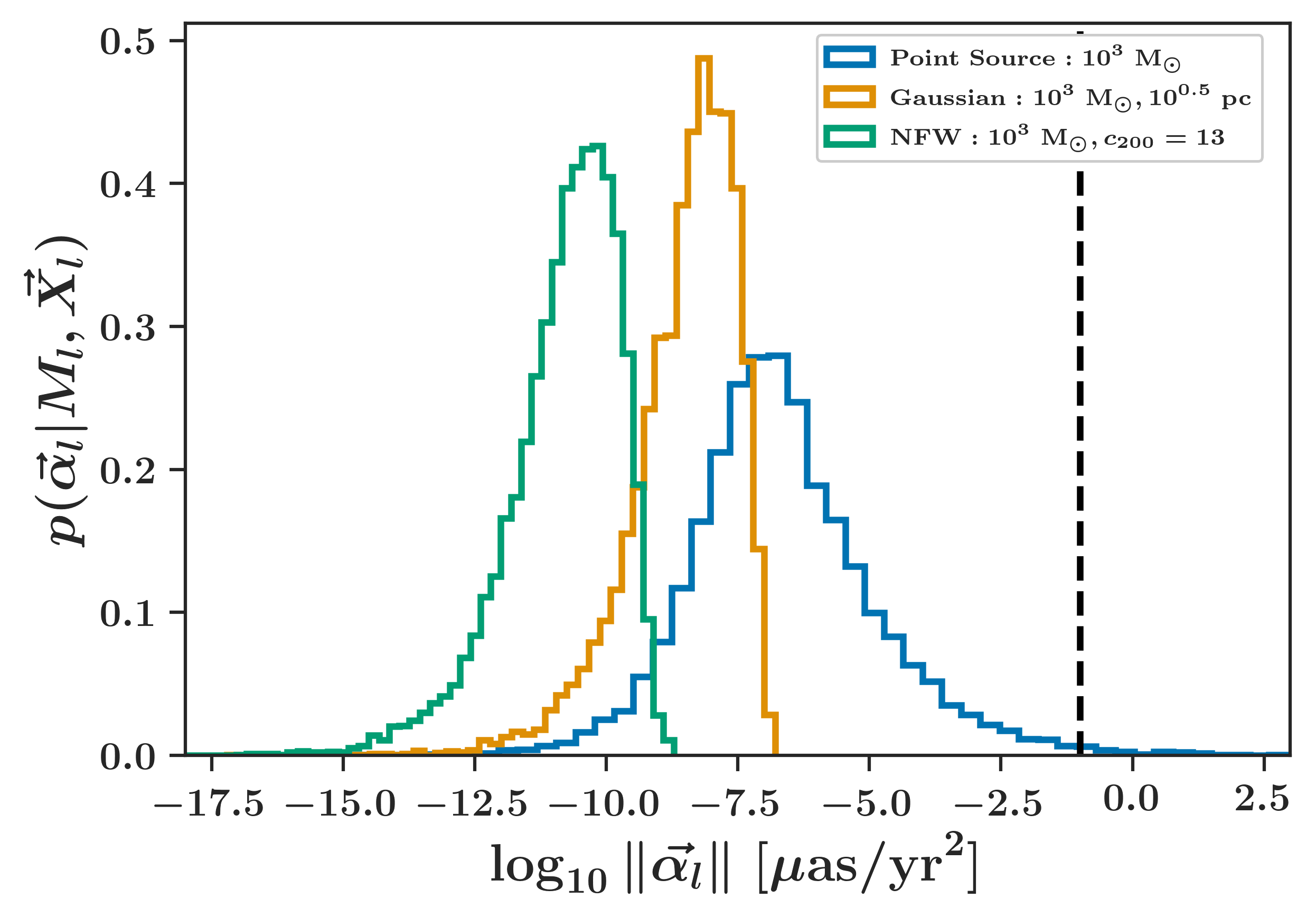}
    \caption{}
    \label{fig:alphal_prior}
    \end{subfigure}
\caption{Probability Distribution Functions for Theory Likelihoods. \textit{a)} Probability distribution functions for the magnitude of the angular size of the impact parameter, $\beta$, of 1 star given a fixed number of lenses, $N_l$. The black, dashed line gives the effective length of one Roman Space Telescope Field of View (FOV; 0.53 deg $=$ 1908 arcsec), assuming a square area with the same area as the true FOV (0.28 deg$^2$). The functions are arbitrarily normalized to fit on this plot. See Appendix A for the derivation of these functions. \textit{b)} Probability of the magnitude of an apparent acceleration for 1 star produced by DM lensing given a specific mass and other lens properties (Equation~\ref{eqn:full_alphal_prior} marginalized over velocity and impact parameter). The different colors correspond to different assumed lens types. The black, dashed line gives the Roman Space Telescope's assumed acceleration sensitivity for 1 star in the EML survey ($\sigma_\alpha = 0.1~\mu\rm{as/yr}^2$).}
\label{fig:priors}
\end{figure*}

The $\beta$ distribution can be found using some geometry and a little bit of numerical integration (see Appendix A for a full derivation). In essence, the calculation is one of finding the probability of the minimum distance between randomly placed points on a plane. It is straightforward to find the probability for the distance between two randomly placed points \citep{Philip2007}. This can then be generalized to $N$ points. Then the minimum can be found using logic. In general, it is a function with a long tail toward low $\beta$ values that peaks near $\beta \sim (2\sqrt{N_l})^{-1} l_{\rm{FOV}}$, where $l_{\rm{FOV}}$ is the side-length of the FOV. Fig.~\ref{fig:impact_prior} shows the form of the function for various numbers of lenses. 

In summary:
\begin{equation}\label{eqn:pb}
    p(\vec{b}) = p(d_l) p(b_{\theta}) p(\beta | N_l) \delta(N_l - \lceil N_l(M_l) \rceil) p(M_l) \; ,
\end{equation}
where $\delta(x)$ is the Dirac Delta function, $\lceil x \rceil$ is the ceiling function, and $N_l(M_l)$ is given by Equation~\ref{eqn-nl}.

The velocity distribution of the lenses is taken to be independent of $M_l$, $N_l$, or $b$. It is set by the potential of the Milky Way, which we set from observations. As with the impact parameter, we assume that the lenses have isotropic velocities, which allows us to set a uniform prior on the angle of the lens velocity on the sky, $p(v_{\theta}) = U[0, 2\pi)$. For the magnitude, $v_l$, we assume the same velocity distribution that Ref.~\cite{Mishra-Sharma2020} assumes:
\begin{equation}\label{eqn:vl}
    p(v_l) = \mathcal{N}(\vec{\mu}_v, \vec{\sigma}_v) \; ,
\end{equation}
where $v_l = \| \vec{v}_l \|$, $\mu_v = 0~\rm{km/s}$ and $\sigma_v = 220~\rm{km/s}$. In addition, we do not allow the lenses to have velocity greater than the escape velocity of the Milky Way, $v < 550~\rm{km/s}$.

Finally, the probability of a star having a particular apparent acceleration, given a lens mass $M_l$, lens velocity $v_l$, and impact parameter $\vec{b}$, is then:
\begin{align}\label{eqn:full_alphal_prior}
    p(\vec{\alpha}_l | M_l, \vec{v}_l, \vec{b})
    &= \frac{1}{4\pi^2} p(v_l) p(d_l) p(\beta | N_l) \nonumber \\
    &\times \delta(N_l - \lceil N_l(M_l) \rceil) p(M_l) \; ,
\end{align}
where the factor in front, $1/4\pi^2$ is given by the priors on $b_{\theta}$ and $v_{\theta}$.

Assuming the measurement errors on the acceleration are Gaussian, it is then straightforward to find the likelihood of a single stellar acceleration measurement given a DM lens with mass $M_l$:
\begin{equation}\label{eq:LM-1star}
    \mathcal{L}(\vec{\alpha} | M_l) = \int p(\vec{\alpha}_l | M_l) \mathcal{N}(\vec{\alpha} - \vec{\alpha}_l, \sigma_{\alpha}^2) d\vec{\alpha}_l\; ,
\end{equation}
where $\vec{\alpha}$ is the observed acceleration, $\mathcal{N}(\mu, \sigma^2)$ is a normal distribution with mean, $\mu$, and variance, $\sigma^2$, $\sigma_{\alpha}$ is the error of the observation, and $p(\vec{\alpha}_l | M_l)$ is given by marginalizing Equation~\ref{eqn:full_alphal_prior} over $\vec{v}_l$, $\beta$, $N_l$, and $d_l$.

For $N_\star$ stars, the full likelihood is then:
\begin{align}\label{eq:LM}
    \mathcal{L}(\vec{\alpha} | M_l) = \prod_{i=1}^{N_{\star}} \Bigg[ \int &d\vec{\alpha}_{l,i} \mathcal{N}(\vec{\alpha}_{i} - \vec{\alpha}_{l,i}, \sigma_{\alpha}^2) \nonumber\\
    &\times \left( \int p(\vec{\alpha}_{l,i} | M_l, \vec{v}_l, \vec{b}) d\vec{v}_l d\vec{b} \right) \Bigg] \; .
\end{align}

It is straightforward to generalize this to lenses that have parameters beyond $M_l$. 

Note that the above likelihood does not explicitly include the possibility of multiple stars being lensed by the same lens. However, some of this signal is preserved -- see Section~\ref{sec:results} for further discussion of this point.

A summary of our likelihood terms is given in Table~\ref{tab:like}.

\begin{table*}
\centering
\begin{tabular}{lcc}
\toprule
Variable & Model Used \\
\midrule
Lens Mass, $p(M_l)$ & Log-Uniform Prior on $[10^0, 10^8]~M_{\odot}$ \\
Fraction of DM in Halos, $p(f)$ & Uniform Prior on $[0,1]$\\
Scale Radius for the Gaussian Profile, $p(R_0)$ & Log-Uniform Prior on $[10^{-4},10^4]~\rm{pc}$\\
Concentration Parameter for the NFW Profile, $p(c_{200})$ & Log-Uniform Prior on $[10^0, 10^8]$\\
Number of Lenses, $p(N_l)$ & Delta Function Set by the Mass of the Lens (Eqn.~\ref{eqn-nl}) \\
Impact Parameter (total), $p(\vec{b})$ & Set by Eqn.~\ref{eqn:pb} and the relevant probabilities below\\
Impact Parameter (on-sky angle), $p(b_{\theta}$) & Uniform prior on $[0,2\pi)$\\
Impact Parameter (distance to lens), $p(d_l)$ & Follows Milky Way DM density, see Eqn.~\ref{eqn:mwdens} \\
Impact Parameter (on-sky angular size), $p(\beta)$ & See Fig.~\ref{fig:impact_prior} and Appendix A \\
Lens Velocity (magnitude), $p(v_l)$ & Normal distribution, see Eqn.~\ref{eqn:vl}\\
Lens Velocity (on-sky angle), $p(v_{\theta})$ & Uniform prior on $[0,2\pi)$\\
\bottomrule
\end{tabular}
\caption{Summary of Likelihood Terms and Assumed Models}
\label{tab:like}
\end{table*}



\subsection{Details of the Sampling}
In this work, we focus on three types of mass profiles for our halos: point source, Gaussian, and NFW. In all cases, we assume a log-uniform prior on the mass within $[10^0, 10^8] M_{\odot}$ and sample in $\log_{10} M_l$.

The point source case is most relevant for MACHOs; however, as we describe in Section~\ref{sec:results}, we expect microlensing to be a better model for these lenses. Thus, it mostly serves as a test case in this work. The point source case has $M(b) = M_l$ for all $b$ and $M_l' = M_l'' = 0$.

Unlike the other two cases, the Gaussian profile is not a realistic profile for any DM models. However, it is a commonly used profile due to its analytical nature. We compute the results here in order to compare to previous work. The profile is defined as:
\begin{equation}
    \rho(r) = \frac{M_0}{(2\pi)^{3/2} R_0^3}\exp\left[-\frac{r^2}{2R_0^3}\right]\; ,
\end{equation}
where $M_0$ is the total lens mass and $R_0$ is a characteristic radius. Here we set $M_0 = M_l$ and we sample in $\log_{10} M_l$ and $\log_{10} R_0$. We use a log-uniform prior of $[10^{-4}, 10^4]$ pc for the scale radius.

The NFW profile, given in Equation~\ref{eqn:mwdens}, is the expected profile for DM. We choose to model the profile using the mass, $M_l$, and concentration parameter, $c_{200}$. Thus we replace $\rho_s$ with:
\begin{equation}
    \rho_s = \frac{200}{3} \frac{\rho_c c_{200}^3}{\ln(x) - c_{200}/x}\; ,
\end{equation}
where $\rho_c$ is the critical density of the universe today, and $x = 1+c_{200}$. The scale radius, $r_s$, is then given by:
\begin{equation}
    r_s = \left(\frac{3M_{200}}{800\pi c_{200}^3 \rho_c}\right)^{1/3} \; ,
\end{equation}
where $M_{200}$ is defined as the mass of a halo with density $200 \rho_c$. Here we set $M_{200} = M_l$ and sample in $\log_{10} M_l$. We also sample in $\log_{10} c_{200}$ with a log-uniform prior of $[10^0, 10^8]$. 

For the EML survey itself, we assume the following: the EML survey will measure the accelerations of $10^8$ stars with sensitivity $0.1~\mu\rm{as/yr}^2$ \citep{Sanderson2019}. We also assume a square FOV\footnote{The true FOV will be a more complicated shape -- see Fig. 2 of Ref.~\cite{Akeson2019}.} with area equal to the true FOV area, $A_{\rm{FOV}} = 0.28~\rm{deg}^2$.

We use the \texttt{emcee} package to perform our MCMC. We only explicitly sample in the mass profile parameters described above. However, each sample of $\log_{10} M_l$ sets $N_l$ via Equation~\ref{eqn-nl}. This then sets the probability distribution for the impact parameter using Equation~\ref{eqn:pb}. We sample $N_\star$ times from this distribution. We then sample $N_\star$ times from the $v_l$ distribution. The apparent acceleration from lensing is calculated using these sampled parameters for each star via Equation~\ref{eqn-alphal}. The likelihood is then calculated via Equation~\ref{eq:LM}. 

\section{Results \& Discussion}\label{sec:results}
Using the forward modeling procedure described above, we forecast the expected limits on DM substructure that the Nancy Grace Roman Space Telescope's EML survey would place from lensing-induced accelerations on stars. Below we describe our results in detail and discuss possible limitations and future work. Our results are summarized in Table~\ref{tab:results}.


\begin{table*}
\centering
\caption{Forecasted 95\% Limits on DM Substructure Using the Roman Space Telescope EML Survey}
\label{tab:results}
\begin{tabular}{l|cccc}
\toprule
      Mass Profile Type & Mass Limits [$\log_{{10}} \rm{M}_{\odot}$] & Radius Limits [$\log_{{10}} \rm{pc}$] & Concentration Limits [$\log_{{10}} c_{200}$] & Fraction Limits \\
\midrule
           Point Source &                             $>0.9~\& <3.4$ &                                    -- &                                           -- &              -- \\
Point Source + Fraction &                             $>0.3~\& <3.2$ &                                    -- &                                           -- &         $<0.89$ \\
               Gaussian &                                    $<7.42$ &                              $>-0.61$ &                                           -- &              -- \\
    Gaussian + Fraction &                                    $<7.45$ &                              $>-0.61$ &                                           -- &         $<0.95$ \\
                    NFW &                                    $<7.49$ &                                    -- &                                      $<2.50$ &              -- \\
         NFW + Fraction &                                    $<7.45$ &                                    -- &                                      $<2.56$ &         $<0.95$ \\
\bottomrule
\end{tabular}
\end{table*}

\begin{figure}
    \centering
    \includegraphics[width=0.5\textwidth]{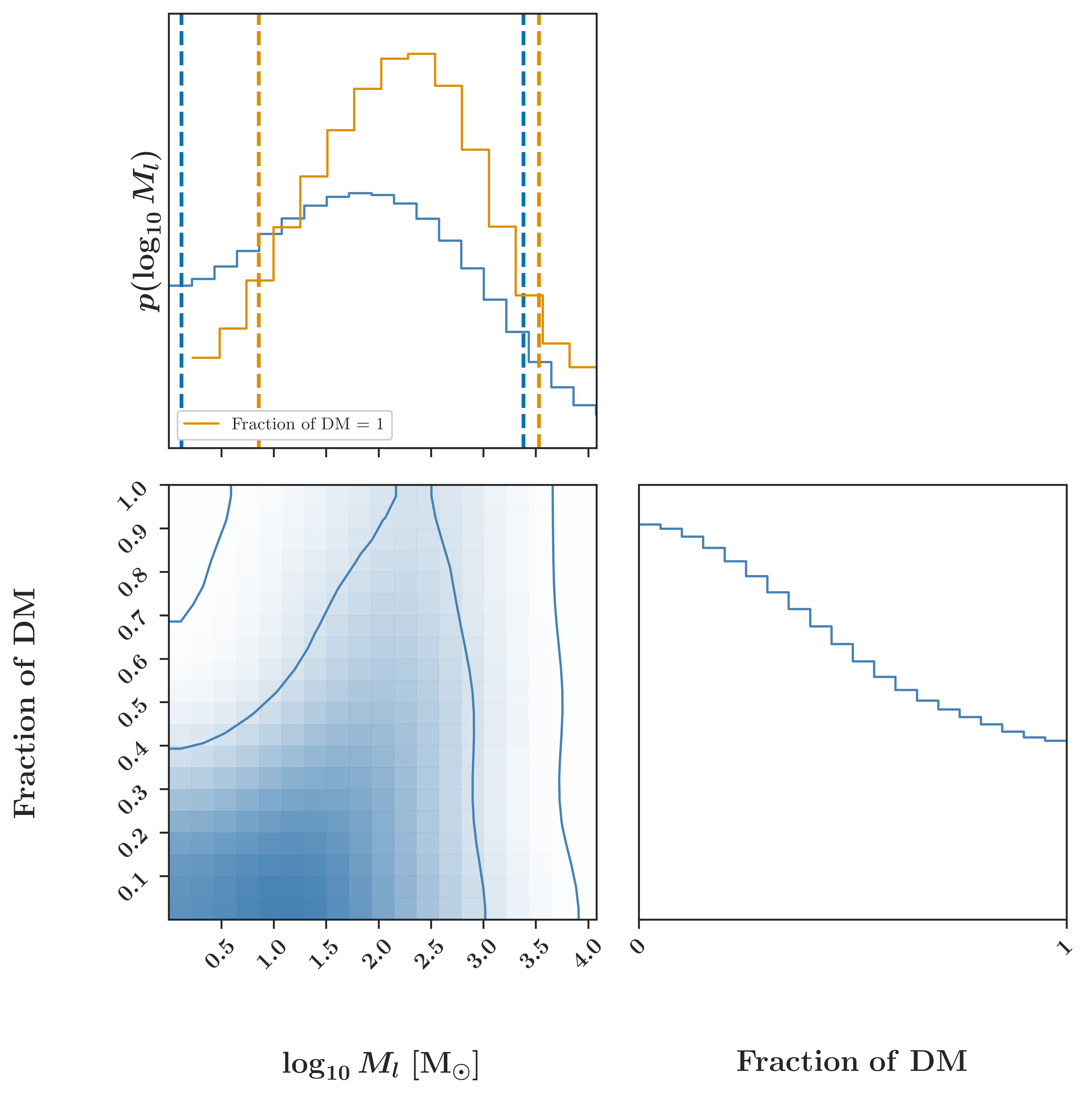}
    \caption{Forecasted Constraints on Point Source Lenses by the Roman Space Telescope's EML Survey. The blue lines/areas show the results when the fraction of DM in these lenses is also sampled. The orange lines give the results when we assume that the fraction of DM in these halos is equal to unity. The dashed, vertical lines give the $95\%$ credible intervals. The 2D plot shows the 68\% and 95\% contours.}
    \label{fig:ps}
\end{figure}

\subsection{Results For Different Lens Profiles}
Here we discuss our results for 3 different types of lenses: point sources, Gaussian profiles, and NFW profiles. For each of these, we consider a case where these halos make up all of the DM and a case where the fraction of DM contained in these halos is treated as a free parameter. 

As a test case, we first consider point source lenses. Note that the effects from these lenses is probably best described with microlensing. However, at large impact parameters they would produce the weak lensing signatures described in this paper. To account for this, we place limits on the maximum apparent acceleration induced by lensing. We limited the maximum acceleration to a signal-to-noise ratio (SNR) of $1$ for any single star. In practice, any stars with acceleration signatures at high SNRs would be cut from the data sample anyway.    

For both the case where we allow the fraction of the DM in these halos to vary and in the nominal case where all of the dark matter is in these subhalos, we find that the Roman Space Telescope will be able to place constraints at both high and low masses. These limits are comparable to previous MACHO constraints \citep{Alcock2001,Tisserand2007, Brandt2016}. The previous MACHO constraints prohibit MACHOs from making up all of the dark matter at any mass. However, the constraints are weakest in the $10-100~\rm{M}_{\odot}$ range. Those constraints allow for MACHOs to make up to 40\% of the DM at 10 M$_\odot$. In the very conservative case we show here, the Roman Space Telescope will place similar constraints using these weak lensing signatures.

Note that the point-source case appears to feature a ``detection", despite the use of fake data with no injected signal (i.e., Gaussian acceleration signals). This ``detection'' is just the weakest point between two limiting cases. At the low-mass end, there are large numbers of lenses, so the probability of a small impact parameter is very large. At the high-mass end, each lens has a large impact on the surrounding stars, since the acceleration signal scales with mass (see Equation~\ref{eqn-alphal}). The peak in the posterior is the maximum mass at which we expect more than $1$ lens within the survey volume. As we describe above, these signals are probably better described via microlensing. The efforts to find planets via microlensing with the Roman Space Telescope \citep{Penny2019, Johnson2020} will also be able to place constraints on MACHOs and other compact objects.

\begin{figure}
    \centering
    \includegraphics[width=0.45\textwidth]{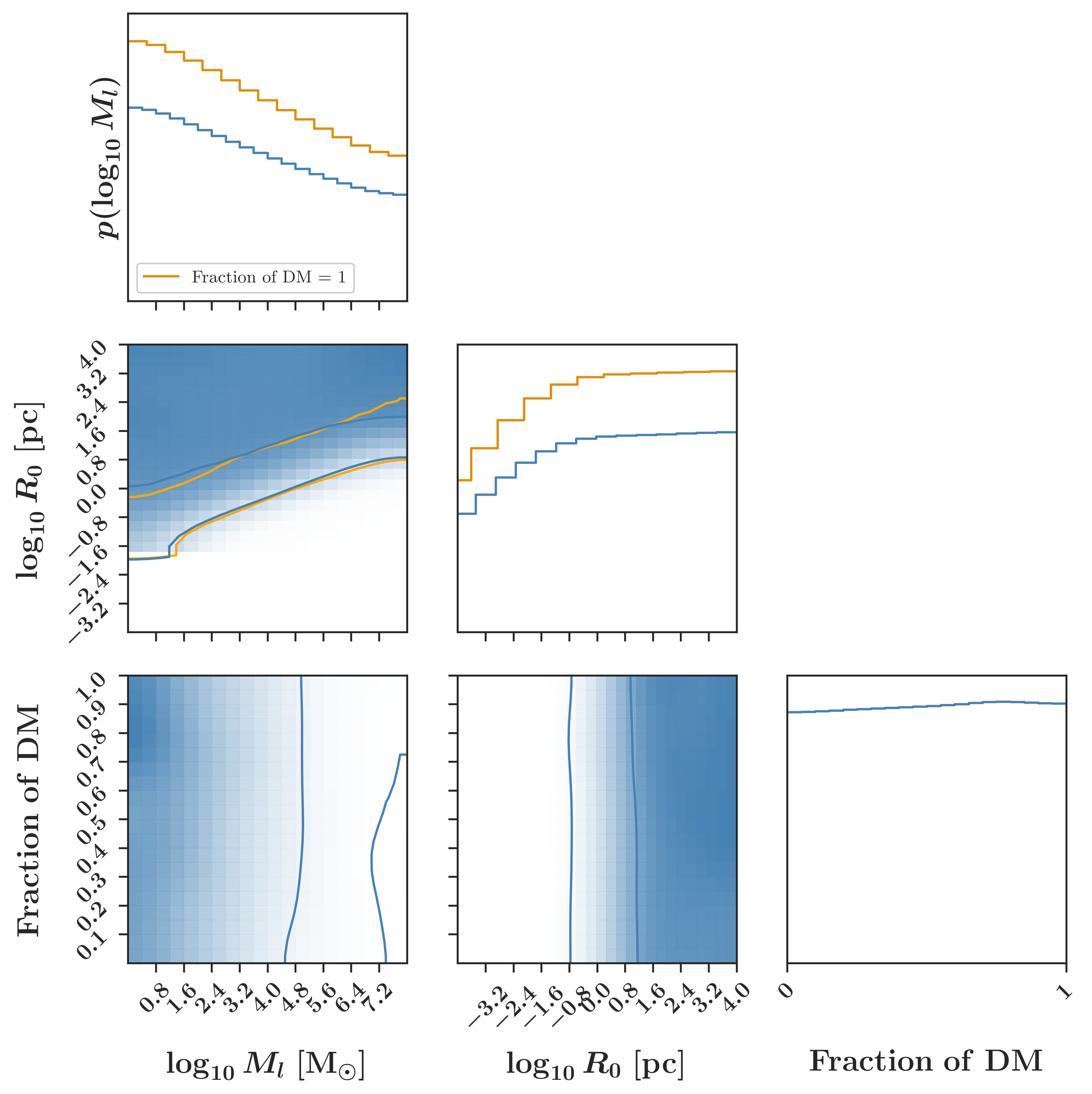}
    \caption{Forecasted Constraints on Gaussian Lenses by the Roman Space Telescope. The blue lines/areas show the results when the fraction of DM in these lenses is also sampled. The orange lines give the results when we assume that all the DM is in these halos. The 2D plot shows the 68\% and 95\% contours.}
    \label{fig:gaussian}
\end{figure}

\begin{figure}
    \centering
    \includegraphics[width=0.45\textwidth]{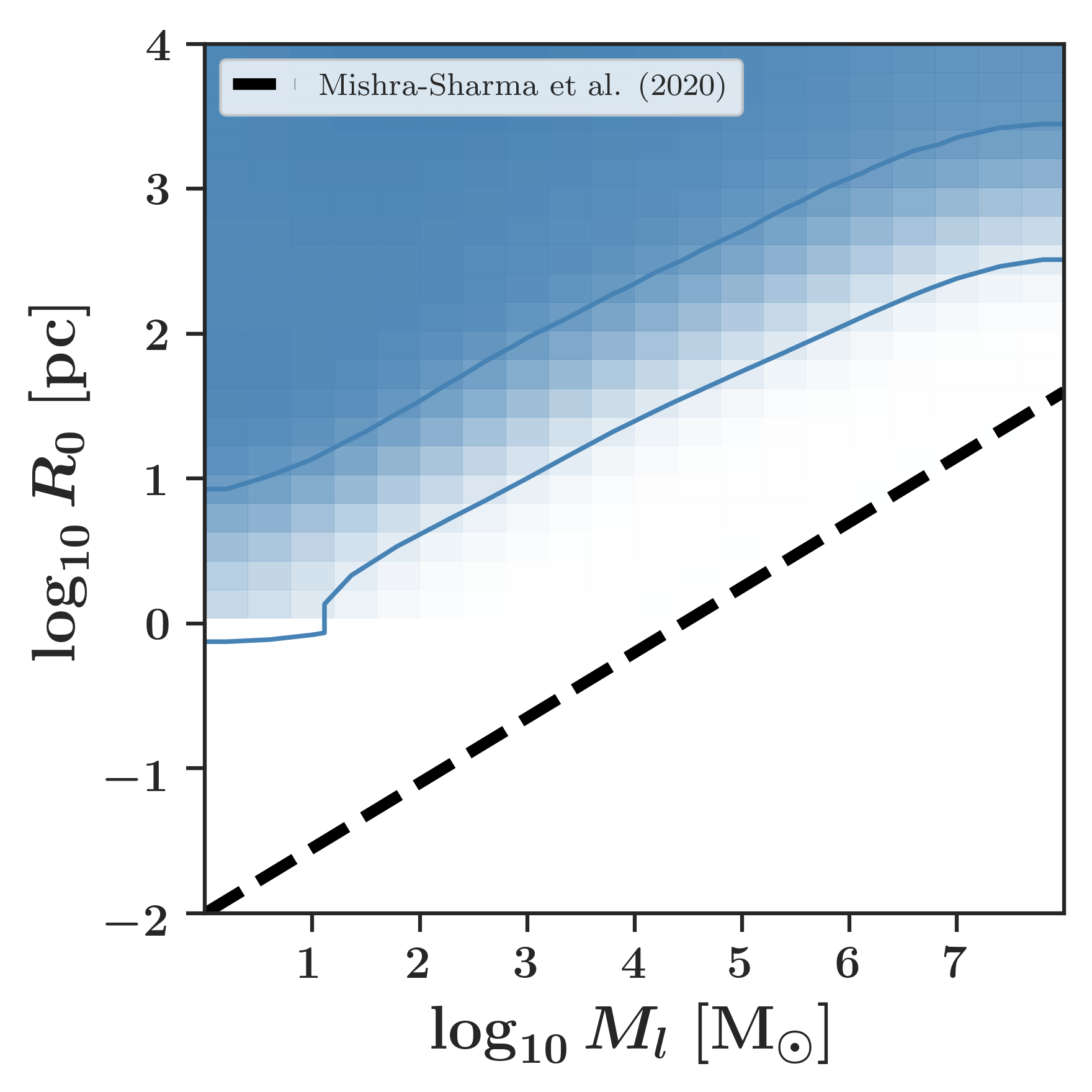}
    \caption{Forecasted Constraints on Gaussian Lenses by a survey like the Roman Space Telescope's EML Survey, but with a larger field of view ($f_{\rm{sky}} = 0.05$). The blue lines give the 68\% and 95\% contours. The dashed black line gives the $95\%$ upper limit from Ref.~\citep{Mishra-Sharma2020}.}
    \label{fig:ms_comp}
\end{figure}

Next, we consider the Gaussian profile case, which is analogous to the ``compact'' lens case that Refs.~\cite{Mishra-Sharma2020, Mondino2020} consider. Fig.~\ref{fig:gaussian} shows the results for the EML survey. In Fig.~\ref{fig:ms_comp} we show the results if we assume the same survey parameters as Ref.~\cite{Mishra-Sharma2020}'s ``WFIRST-Like'' survey (similar to the EML, but with a wider field of view). We find somewhat tighter constraints than the results quoted in Ref.~\cite{Mishra-Sharma2020}. This is most likely due to the use of forward modeling rather than the power spectrum approach. The acceleration signature is not Gaussian, so the power spectrum approach misses part of the signal. Our result suggests that an alternative approach using higher order statistics, such as a bispectrum or trispectrum analysis, would be fruitful. 

\begin{figure*}
    \centering
    \includegraphics[width=0.95\textwidth]{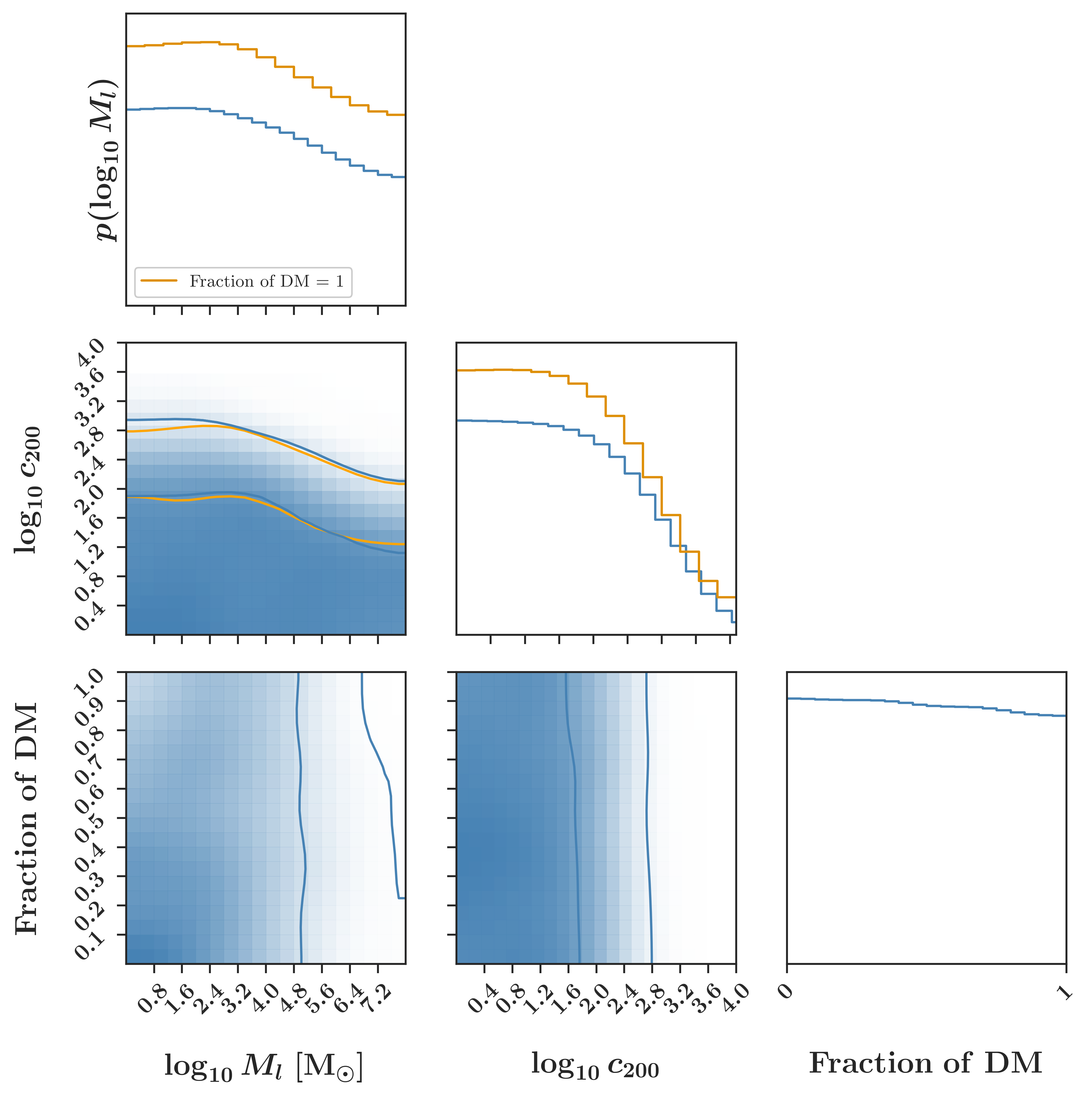}
    \caption{Forecasted Constraints on NFW lenses by the Roman Space Telescope. The blue lines/areas show the results when the fraction of DM in these lenses is also sampled. The orange lines give the results when we assume that all the DM is in these halos. The 2D plot shows the 68\% and 95\% contours.}
    \label{fig:nfw}
\end{figure*}

Finally, we consider the NFW profile case (see Fig.~\ref{fig:nfw}). The mass constraints for this case are the least constraining, as is expected given the diffuse nature of these halos. However, we find that the Roman Space Telescope would place interesting constraints on the concentration of these halos. Namely, it would be very sensitive to any halos with $c_{200} > 10^4$. It would even be able to probe down to $c_{200} \sim 100$ at a $2\sigma$ level for $M\sim 10^8~\rm{M}_{\odot}$. For typical NFW halos in CDM, we expect $c_{200}$ values of around $10-20$ for Milky Way-sized objects. However, for subhalos very close to their host galaxies, simulations have found that $c_{200}$ could extend upwards to $\sim 70$ \citep{Moline2017}. Thus, with only a modest improvement in sensitivity (either through analysis methods or survey strategy), the Roman Space Telescope's EML survey could be sensitive to CDM subhalos within the galaxy.

\subsection{Discussion \& Future Work}

We now discuss the caveats of this study. First and foremost, the EML survey parameters have not been finalized and are subject to change. As mentioned above, we also make several simplifications (e.g., shape of the FOV) that could marginally effect the results. We discuss the most impactful assumptions below: neglecting binary star systematics, the use of mono-mass populations of lenses, and our modeling of multiple star lensing by single lenses.

One possible systematic is the acceleration from binary star systems. These systems are ubiquitous and could be a serious systematic since they would produce large acceleration signatures. For low-mass binary systems, the most likely semi-major axis is $a=5.3~\rm{AU}$ \citep[for a review of binary star systems, see Ref.][]{Duchene2013}. This, with $M=0.1~M_{\odot}$ and a distance of $8~\rm{kpc}$ from the observer gives an on-sky acceleration of $\alpha \sim 35~\mu\rm{as/yr}^2$, which is many orders of magnitude greater than the expected lensing signal from a population of lenses. However, careful modeling could be used to distinguish the signals. In most cases, the periodic motion of the stellar positions would allow us to discard these sources \citep[see][for a similar discussion of how to distinguish binaries from planet signals]{Penny2019}. 

We only consider mono-mass populations of lenses in our work. This allows us to consider the constraining power at different masses more easily, but it is not a realistic scenario. The expected subhalo mass function is largely uncertain, due to tidal effects, baryonic feedback, and DM physics. In addition, the radial dependence and expected profiles of the subhalos can also heavily depend on these factors and vary from the assumptions made in this paper \citep[see, e.g.,][]{Garrison-Kimmel2017, Green2021, Nadler2021}. Future work could look into the constraints on the mass function, as well as consider other radial dependencies and mass profiles. Constraints on these properties would let us learn about both DM physics and baryonic feedback.

Finally, as mentioned in Section~\ref{sec:methods}, we do not explicitly consider the effects of a single lens affecting multiple stars. This would add extra correlations to the acceleration signal expected in the stars. In some sense, our $p(\beta | N_l)$ likelihood does account for the particular distribution of impact parameters assuming a fixed number of lenses in the field. The magnitude of the signal should be preserved by our formalism, even if the angular distribution is not. Our analysis here can be taken as a conservative estimate of the true signal. 

There are also several ways of improving these constraints. We discuss 2 avenues for further improvement here: 1) using time domain signatures; 2) using observations of other systems to improve constraints.

This paper solely considered the overall acceleration measurements of these stars at the end of the Roman Space Telescope's nominal mission. However, we could also consider the information contained in the time series measurements of these sources. For example, consider a lens 1 kpc from us moving at 300 $\rm{km/s}$ in front of a background of $10^8$ stars distributed across the 2 deg$^2$ EML survey area. If this lens was directly in front of one star, it would take 60 years to be in the line of sight of a different star. Over 5 years, we still would expect to see a large difference in the lensing pattern from this lens, since the signal scales so strongly with impact parameter. A detailed exploration of this topic is left to future work.

One other way of improving constraints is to use complementary observations to the EML survey. A guest observer program for the Nancy Grace Roman Space Telescope that targets nearby galaxies could yield even better limits, provided the number of stars observed at high astrometric accuracy is sufficient. For example, an M31 survey combined with the Hubble Space Telescope PHAT survey \citep{Dalcanton2012} could be incredibly powerful. This would give a longer baseline to the observations and allow us to leverage the time domain signatures we discuss above. However, the EML survey's timing resolution will likely allow it to remain the best survey for observing these lenses.

In this paper, we considered the constraints the Nancy Grace Roman Space Telescope's Exoplanet Microlensing (EML) Survey would place on dark matter substructure through their lensing effects on the apparent accelerations of stars in the Galaxy. We forward modeled the effects of various lens profiles using a semi-analytic framework. While Roman will not be able to place strong constraints on the fraction of DM in these lenses, we found that Roman's EML survey will place competitive bounds on the masses and other profile parameters of point source, Gaussian, and NFW lenses. In particular, the NFW bounds will be complementary to those placed by satellite counts by probing lower masses than is possible with MW satellites \citep[see, for example,][]{Drlica-Wagner2019}. Future work on the timing signatures of these lenses could lead to further improvement on these bounds.

\acknowledgements
 The authors would like to thank Siddharth Mishra-Sharma for his very helpful comments. This work was done at the Jet Propulsion Laboratory, California Institute of Technology, under a contract with the National Aeronautics and Space Ad-ministration. This work was supported by NASA grant 15-WFIRST15-0008 \textit{Cosmology with the High Latitude Survey} Roman Science Investigation Team (SIT). \copyright~2021. California Institute of Technology. Government sponsorship acknowledged.

 \textit{Software:} astropy \cite{astropy}, emcee \cite{Foreman-Mackey2013}, matplotlib \cite{matplotlib}, numpy \cite{numpy}, scipy \cite{scipy}

\appendix

\section{Derivation of the Impact Parameter Prior}

We would like to find the distribution of impact parameters given a random set of DM lenses and stars. First, note that the number of stars does not matter for this distribution -- the populations of each are totally separate, and we can consider each star independently. Essentially, each star is a random draw of the impact parameter distribution.

Throughout this appendix, we keep the lenses at a fixed distance from us and only consider their on-sky distribution. The distance and on-sky distributions should be totally independent and the distance can be sampled afterwards.

First consider a toy problem: a 1D line with length $a$ where we place uniformly place our lenses. For now, we place a single lens. Then place a star on this line. It too will be randomly and uniformly placed. The probability of the squared distance between the star and the lens, $x$, is given by \citep{Philip2007}:
\begin{equation}
    f(x) = \frac{1}{a\sqrt{x}} - \frac{1}{a^2} \;, \; \mathrm{for} \; 0\leq x\leq a^2 \; ,
\end{equation}
and $0$ elsewhere.

Consider $N_l$ lenses. Let the random variable $X_i$ for $i \in [1, N_l]$ be a random variable describing the squared distance of lens $i$ from our one star. As Eq.~\ref{eqn-alphal} shows, the acceleration signal depends very strongly on the impact parameter ($\alpha \propto 1/b^3$). The signal is then mostly determined by the lens closest to the star. Thus, we want to find the probability that the \textit{closest} lens has a squared distance, $x$. In other words, we want: $p(\mathrm{min}(X_1, X_2,..., X_n) = x)$.

Let us start by working with the cumulative distribution function (CDF), $P(\mathrm{min}(X_1, X_2,..., X_n) \leq x)$. Note that this is equivalent to the probability of \textit{any one} lens having squared distance less than $x$. The probability of this is then the converse of the probability that all of the lenses have squared distances greater than $x$. The probability that any one lens has a squared distance greater than $x$ is: $P(\mathrm{any}(X_1, X_2,..., X_n) \geq x) = 1-F(x)$. For all $n$ lenses this is: $P(\mathrm{all}(X_1, X_2,..., X_n) \geq x) = (1-F(x))^n$. Finally, the CDF we are looking for:
\begin{equation}\label{eqn:cdftomin}
P(\mathrm{min}(X_1, X_2,..., X_n) \leq x) = 1 - (1-F(x))^n \; .
\end{equation}

To obtain the probability distribution function (PDF) in this case, we simply take the derivative of Eqn.~\ref{eqn:cdftomin}. This gives:
\begin{equation}\label{eqn:pdf1d}
    p(\mathrm{min}(X_1, X_2,..., X_n) = x) = \frac{an}{\sqrt{x}-ax} \left( a\sqrt{x} - 1\right)^{2n}
\end{equation}

Now, consider the 2D case. To be totally general, we can consider a rectangle with sides $a$,$b$, and $a\leq b$. For just 1 star and 1 lens, the probability of a given squared distance is \citep{Philip2007}:
\begin{equation} \label{eqn:phillips2D}
  g(x) =
    \begin{cases}
      -2\frac{\sqrt{x}}{a^2b} -2\frac{\sqrt{x}}{ab^2} + \frac{\pi}{ab}\\ + \frac{x}{a^2b^2} \; ,

      & \text{if $0\leq x \leq a^2$} \; ;\\

     -2\frac{\sqrt{x}}{a^2b} -\frac{1}{b^2} \\
     + \frac{2}{ab}\arcsin{\left( \frac{a}{\sqrt{x}} \right)}\\
     + \frac{2}{a^2b}\sqrt{x-a^2}\; ,

      & \text{if $a^2\leq x \leq b^2$} \; ;\\

     -\frac{1}{b^2} + \frac{2}{ab}\arcsin{\left( \frac{a}{\sqrt{x}} \right)}\\
      + \frac{2}{a^2b}\sqrt{x-a^2} -\frac{1}{a^2} \\
      + \frac{2}{ab}\arcsin{\left( \frac{b}{\sqrt{x}} \right)} \\
      + \frac{2}{ab^2}\sqrt{x-b^2}\\
      - \frac{\pi}{ab}-\frac{x}{a^2b^2} \; ,

      & \text{if $b^2\leq x \leq a^2+b^2$} \; ,
    \end{cases}
\end{equation}
and $0$ elsewhere. The generalization to the minimum case is the same as in the 1D case. We numerically integrate Eqn.~\ref{eqn:phillips2D} to obtain it's CDF, G(x). We then apply Eqn.~\ref{eqn:cdftomin} and take the derivative to obtain the PDF.

\bibliography{ref}

\end{document}